# POWER OPTIMIZED PROGRAMMABLE EMBEDDED CONTROLLER

M.Kamaraju[1]        Dr.K.Lal Kishore[2]        Dr.A.V.N.Tilak[3]

[1]Dept. of ECE, Gudlavalleru Engineering College, Gudlavalleru - India
madduraju@yahoo.com
[2]Director,SIT, Jawarlal Nehru Technological University ,Hyderabad- India
lalkishorek@yahoo.com

[3]Dept. of ECE,Gudlavalleru Engineering College, Gudlavalleru - India
avntilak@yahoo.com

*ABSTRACT*

*Now a days, power has become a primary consideration in hardware design, and is critical in computer systems especially for portable devices with high performance and more functionality. Clock-gating is the most common technique used for reducing processor's power. In this work clock gating technique is applied to optimize the power of fully programmable Embedded Controller (PEC) employing RISC architecture. The CPU designed supports i) smart instruction set, ii) I/O port, UART iii) on-chip clocking to provide a range of frequencies  , iv) RISC as well as controller concepts. The whole design is captured using VHDL  and is implemented on FPGA chip using Xilinx .The architecture and clock gating technique together is found to  reduce the power consumption by 33.33% of total power consumed by this chip.*

## 1. INTRODUCTION

Low power consumption in embedded systems [1] has become a key factor for many applications. Portable applications,  needing  long  battery  life  together  with  high peak- performance, are demanding a very careful design at all levels.

The most important factor contributing to the energy consumption is the switching activity [2] [3] .Once the technology and supply voltage have been set , major energy savings come from the careful minimization of switching activity.

While some switching activity is functional, i.e. it is required to propagate and manipulate information; there is a substantial amount of unnecessary activity in almost all digital circuits. Unnecessary switching activity arises from spurious transitions [4] due to unequal propagation delays  (glitches)  and  transitions  occurring  within  units  that  are  not  participating  in  a computation. One way to avoid these activities is by dynamically turning off the clock[5] to unused logic or peripherals .

Existing microprocessors [6], especially in the category of microcontrollers, often have the capability of partially gating the clock signal when they fetch and execute  NOP  or other minimal-activity instructions and also when a peripheral is powered-down. This does not totally eliminate spurious power consumption. Other processors [7] can enter low-power operating modes, disabling the clock signal generator (Xtal oscillator and PLL). However, this is not totally efficient in important applications. In [8] optimization of power is achieved by taking assistance of compiler while executing instructions. However hardware approach proposed in this work reduces the power without compromising speed performance of the chip.

In the proposed PEC, the control unit is designed to have the capability of gating the clock signal when they fetch and execute instructions. On-chip clocking [9] mechanism is employed to synchronize with on-chip peripherals/memory and with the external bus. For power optimization of integrated circuits it is relevant to understand the causes of power dissipation. Clock power[10] dominates the total power consumed by a microcontroller as the clock is fed to most of the circuit blocks. Charge/discharge power given by $P = f C_L V_{dd} V_s$ dominates the total power dissipation of the chip. The frequency f of the clock cannot be reduced as it effects the speed of the chip. When output swings from 0 to $V_{dd}$ then P varies as square $V_{dd}$ .However lowering P by reducing power supply voltage to 2V or less is found to lead to several problems[11] [12] like decrease in drivability of MOSFET and increase in gate delay time.

## 2. PROGRAMMABLE EMBEDDED CONTROLLER ARCHITECTURE

Architecture of PEC is shown in Figure.1. Various blocks in the architecture are register file, ALU, RAM, ROM, UART, I/O Ports, BCD to 7 segment driver , Control unit, and clocker, designed to perform particular task. Register File is a set of registers that are modeled as RAM of 16 bit words, used to store intermediate values during instruction processing. The ALU performs 16 bit operations. The Read Only Memory (ROM) is 256 bytes with 16 bit word length and is used to store the instruction data. The Random Access Memory (RAM) 1K×16 is used to store temporary data. Port 0 and Port 1 are two ports which are configured as output and input ports respectively.

 A display driver for BCD to 7 segment display is designed to drive the 7 segment display unit. The control unit generates various control signals to all other blocks to execute desired task specified by the instructions. The PEC is initiated by the reset signal whenever reset signal asserts high, the controller generate appropriate signals to load the PC address of the ROM. The external interrupt mechanism activates on any hardware interrupt or reset signal arriving at the controller when it is in idle mode.

### 2.1. On-Chip Clocking Mechanism

The frequency of the application specific hardwired oscillator shown in Figure.2 is programmable by means of the 4-bit number (control word value) contained in the dedicated register r_osc. On-chip clocking is used to obtain different frequencies ranging from 44 MHz to 134 MHz by changing the control word values as shown in Figure. 3.

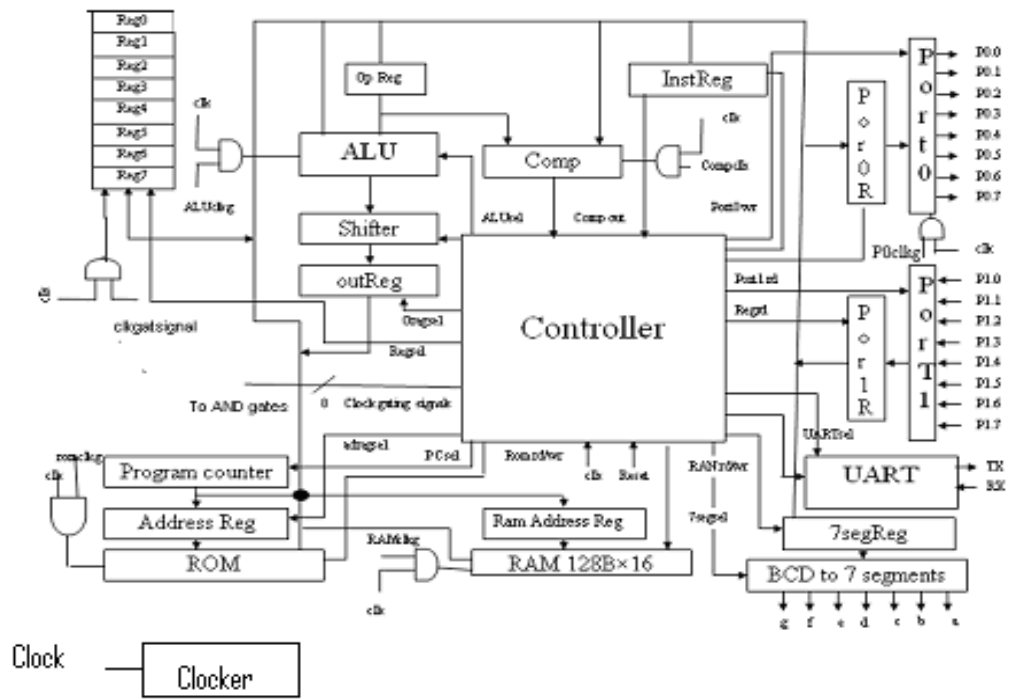

Figure.1 Architecture of Programmable Embedded Controller

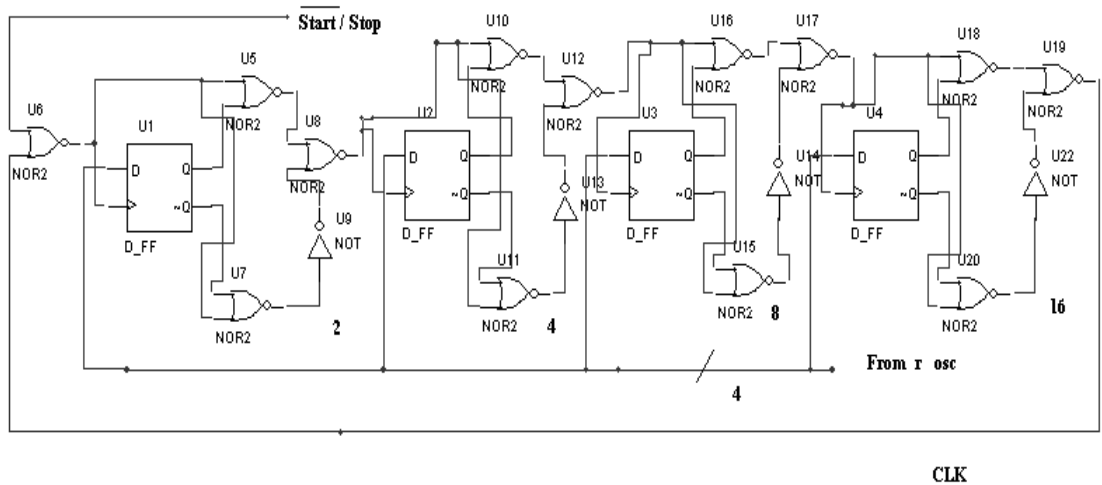

Figure.2 Oscillator Circuit

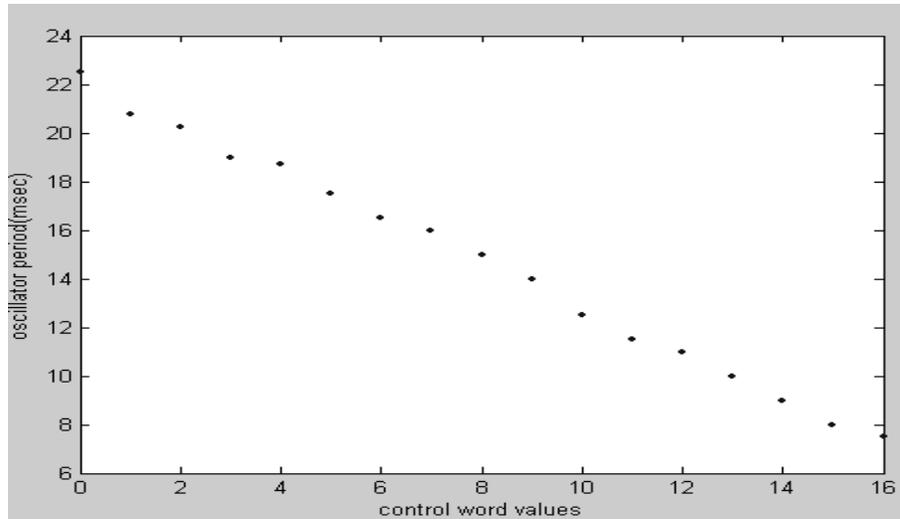

Figure.3 Oscillator Cycle Time vs. Control Word Value

## 2.2. Clock Gating

Figure. 4(a) shows the schematic of latch element. A significant amount of power is consumed during charge/discharge cycle of the cumulative gate capacitance $C_g$ of the latch, when the clock is fed directly and there is no change in the clock cycle[13]. Figure. 4(b) shows the latch with gated clock. By gating the clock [14] [15] , charge/discharge of $C_g$ can be effected only when there is change in the clock cycle thus saving power.

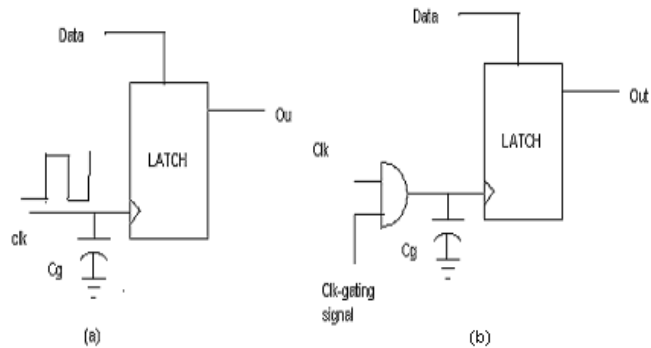

Figure.4 Schematic of Latch Element (a) without clock gating (b) with clock gating

The controller shown in Figure.1 supports predefined smart instruction set having length of 16 bits each. Gated clock signal generated by the control unit allows the clock to be  fed  only to the active blocks and not to the unused blocks.

## 2.3. Control Unit

The control unit provides all of control signals to regulate the data traffic and necessary signals to perform the desired functions. The control unit architecture contain a state machine that causes all appropriate signal values to update based on current state and input signals and produce a next state for state machine. The control unit performs two processes. The first is a combinational process (not clocked) that examines the current state and all inputs and produces output control signals and next state output. The second is the sequential process (having a clock) that is used to store the current state and copy of the next state to the current state.

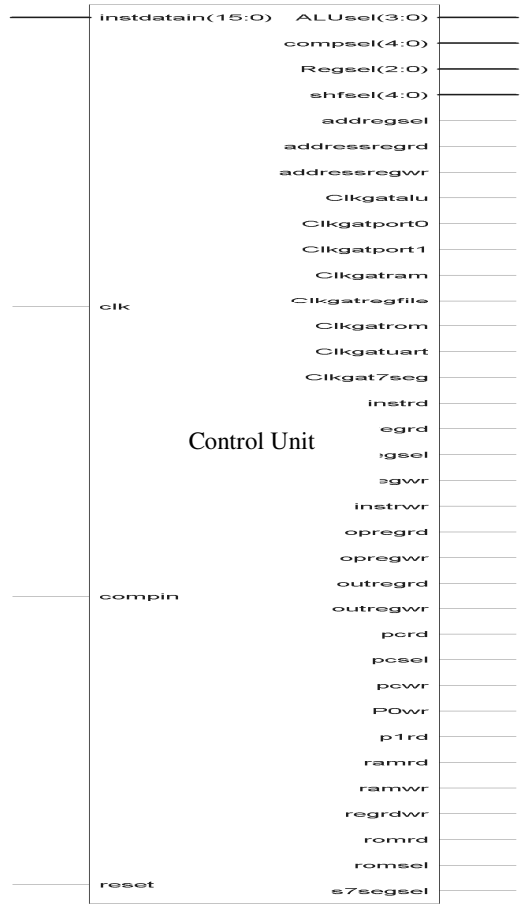

Figure.5 Signals of Control Unit

If the reset signal is high the sequential process set the current state value to reset1, the first state of the reset sequence. The logic for clock gating is implemented within the control unit. The controller generate appropriate clock gating signal to reduce power consumption of the chip. When the control unit decodes the opcode of the instruction, the control unit generates control signals as shown in Figure.5, to execute the instruction.

By implementing the instructions given in the instruction set (Table 1) does not cause any functional limitation, but enables an effective way of power saving through generation of gated clock signals. All the instructions are of length 2 bytes and of direct addressing mode type. Instruction set includes Load, Store, Branch, ALU and Shift instructions.

Table 1  Instruction Set

| OPCODE | INSTRUCTION | DESCRIPTION |
|--------|-------------|-------------|
| 00000 | NOP | No operation |
| 00001 | LOAD | Load register |
| 00010 | STORE | Store the register |
| 00011 | MOVE | Move the value into the register |
| 00100 | LOADI | Load the register with immediate value |
| 00101 | BI | Branch to immediate address |
| 00110 | BGTI | Branch greater than to immediate address |
| 00111 | INC | Increment |
| 01000 | DEC | Decrement |
| 01001 | AND | Logical AND two registers |
| 01010 | OR | Logical OR two register |
| 01011 | XOR | Logical XOR two register |
| 01100 | NOT | Logical NOT the register |
| 01101 | ADD | Add two registers |
| 01110 | SUB | Subtract two registers |
| 01111 | ZERO | Zero a register |
| 10000 | PORT0 | Port 0 write |
| 10001 | BLT | Branch lass than |
| 10010 | BNEQ | Branch not equal |
| 10011 | PORT1 | Port 1 read |
| 10100 | BGT | Branch greater than |
| 10110 | BCH | Branch all the time |
| 10111 | BEQ | Branch if equal |
| 11000 | B7S | 7 segment driver |
| 11001 | BLTE | Branch less than or equal |
| 11010 | SHL | Shift left |
| 11011 | SHR | Shift right |
| 11100 | ROR | Rotate right |
| 11101 | ROL | Rotate lrft |
| 11110 | UARTS | UART sel. |

## 3. RESULTS

The entire design is captured in VHDL and simulated using Xilinx tool. The simulation results of control unit are presented in Figure.6.

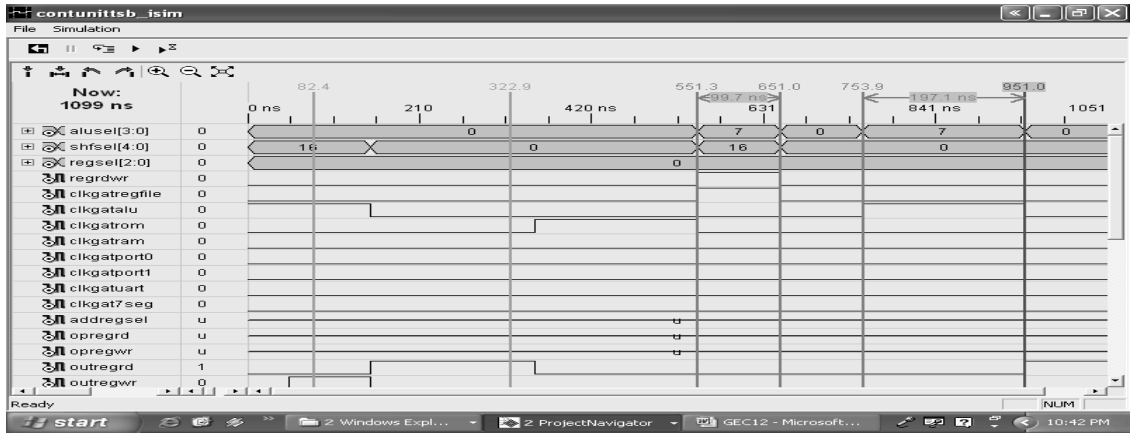

Figure.6 Control Unit Simulation Results

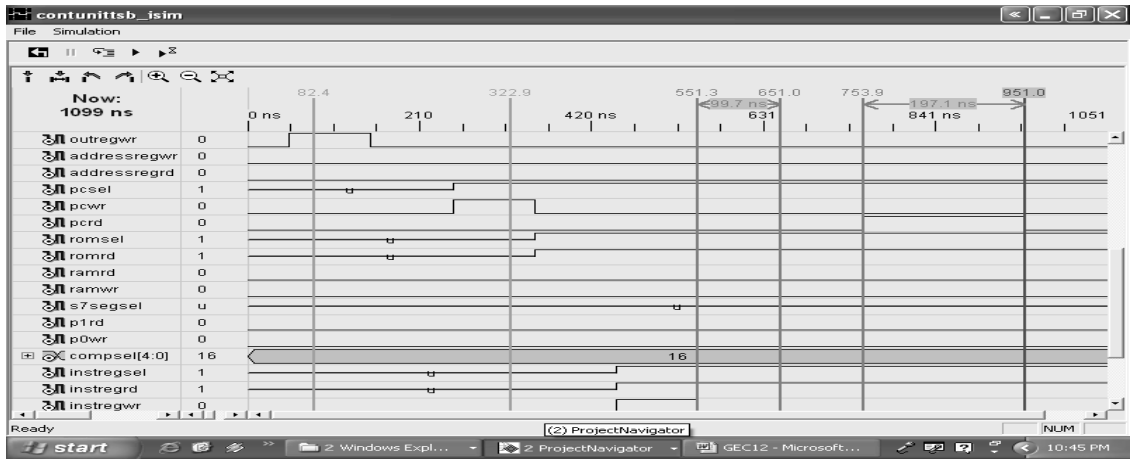

Figure.6 Control Unit Simulation Results (contd…)

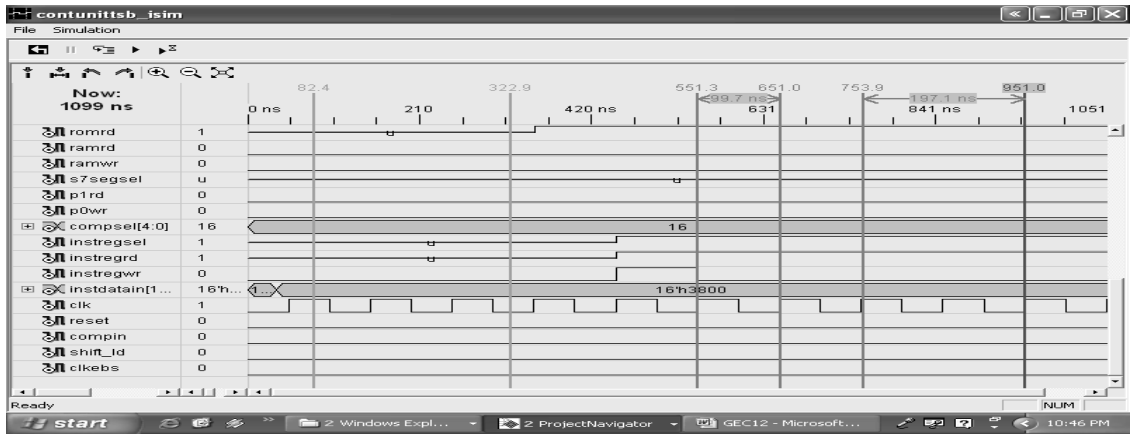

Figure.6 Control Unit Simulation Results (contd…)

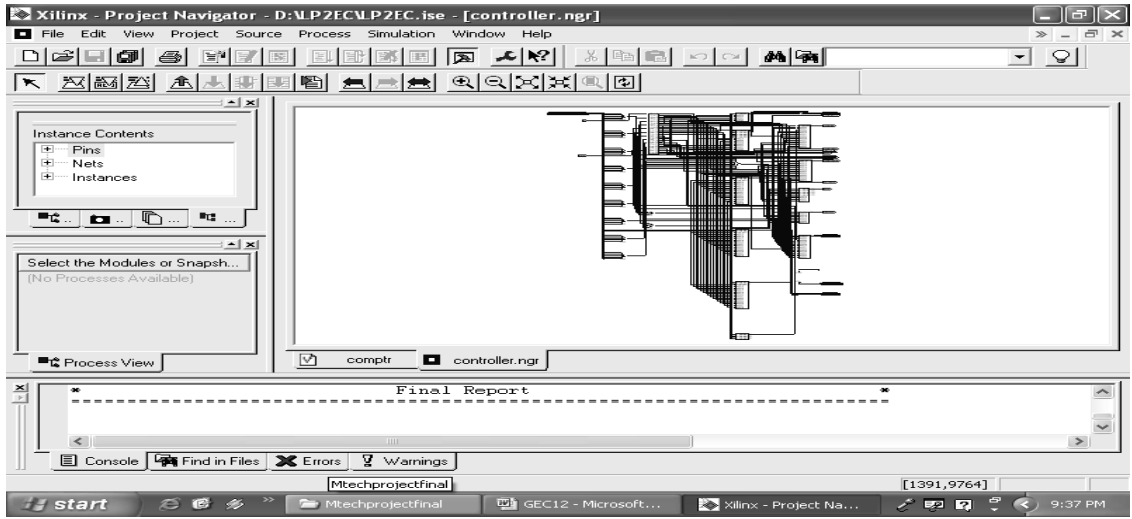

Figure.7 Control Unit RTL Schematic

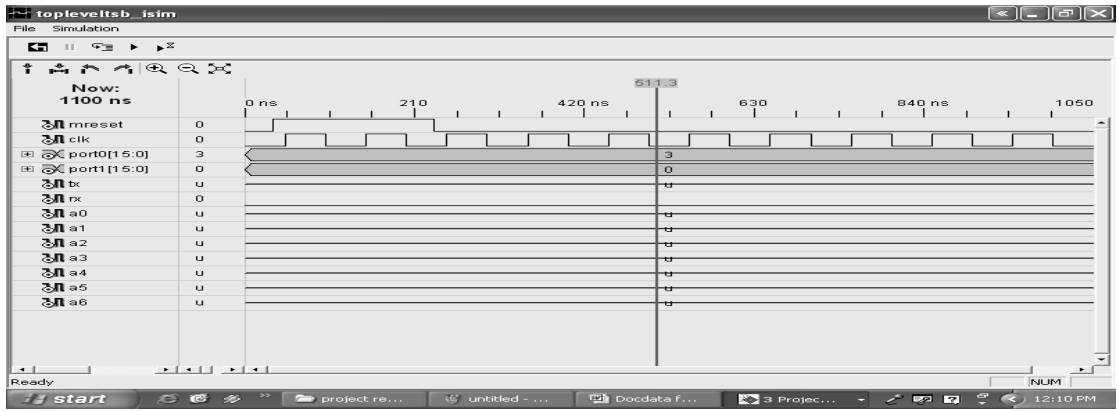

Figure.8 Top Level Module Simulation Results

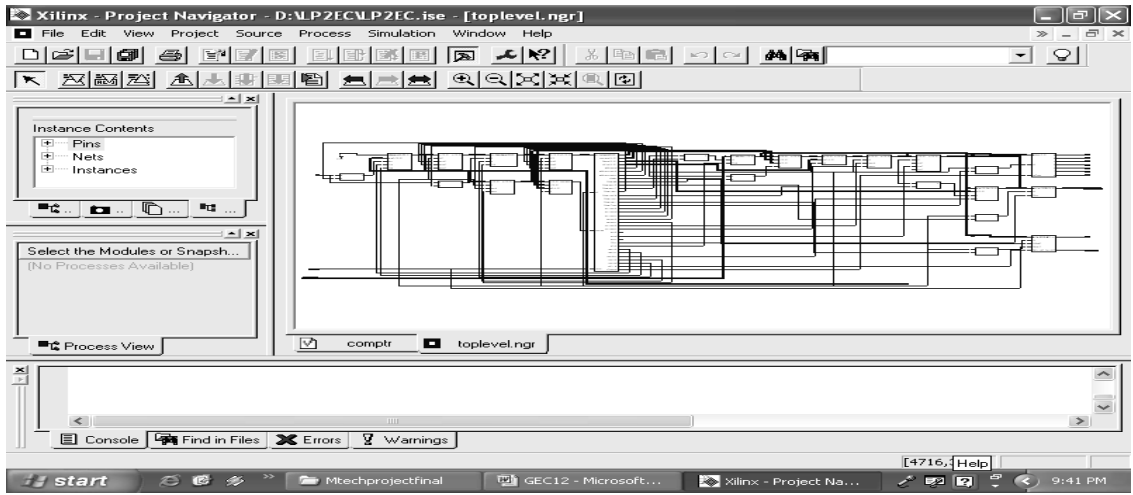

Figure.9 Top Level Module RTL Schematic

### 3.1. Power Analysis

The estimation of power consumption of each module is done using Xilinx Xpower's tool. The graphical representation of power consumed in various modules is shown in Figure.10.

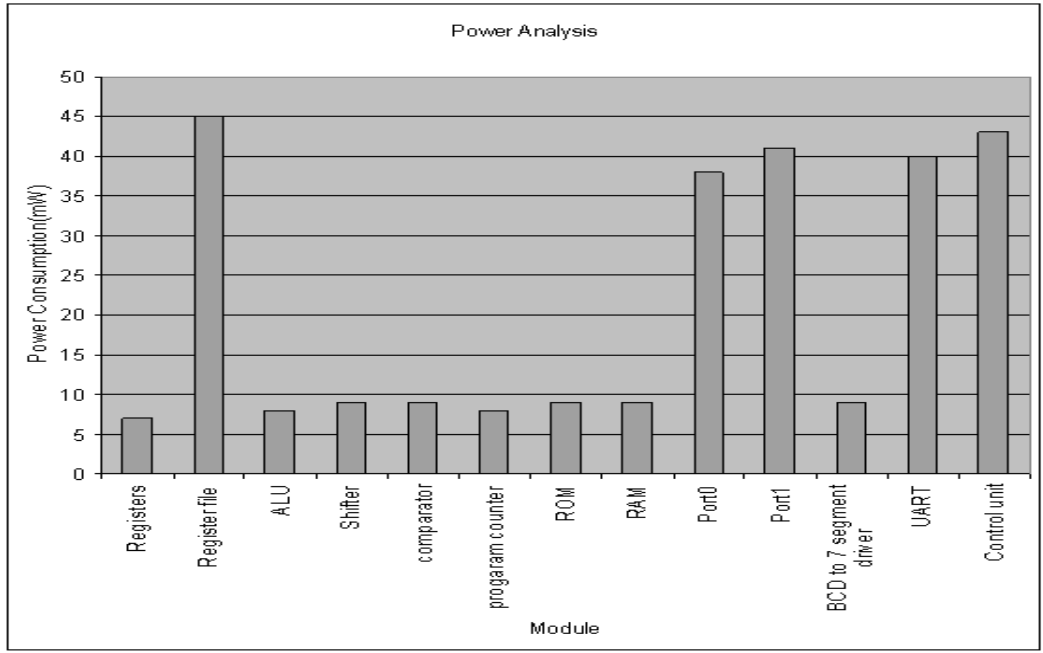

Figure.10 Power Analysis

Total power consumption is estimated to be 273mW without clock gating and only 182 mW after clock gating technique is employed, thus achieving a power saving of 33.33% (Figure.11).

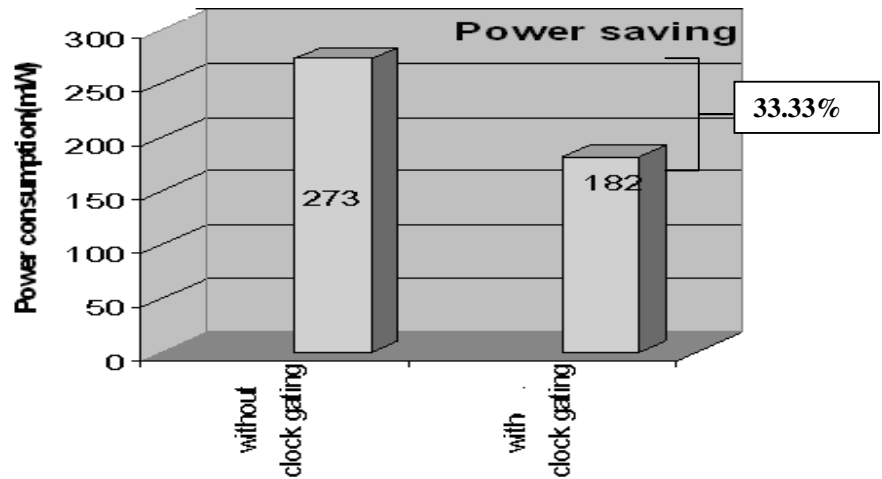

Figure.11 Comparison of Power Consumed without and with Clock Gating

### 3.2. Characteristics of the Chip

The characteristics of the designed chip are

| | |
|---|---|
| Architecture | RISC |
| Optimization | Power |
| Instructions | 2 byte |
| ROM | 256 bytes |
| RAM | 1 KB |
| ALU | 16 bit |
| Power Supply | 2.4V |
| Power Dissipation | 3.62mW/MHz |

## 4. CONCLUSIONS

The need for low power systems is being driven by many market segments. There are several approaches to reducing the power. In this work clock gating technique is applied to optimize the power of fully programmable embedded controller employing RISC architecture. The whole design is captured using VHDL language and is implemented on FPGA chip using Xilinx .The chip has less hardware complexity as this works based on single addressing mode to access the data for processing. The architecture and clock gating technique together have reduced the power consumption by 33.33% of total power consumed by the chip. This clock gating technique can be applied from chip level to module and then eventually to systems.

**Authors**

Dr.K.Lal Kishore , obtained Master's degree and Ph.D. from  Indian  Institute of Science  (IISC) Bangalore.He had published more than 76 research papers in International/Natioanl  journals and presented papers in International / National Conferences.He is best teacher awardee from the Government  of Andhrapardesh,S.V.C.Aiya Memorial Award from IETE,Merit Award from DEC, Ethopia, Bapu Seetharam Memorial Award from IETE  and has many other academic distinctions to his credit.He wrote books on Electronic Devices,Circuit Analysis,Linear IC Applications,Electronic Measurements and Instrumentation and VLSI Design.

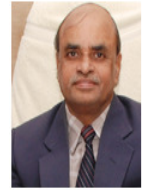

Dr.A.V.N.Tilak, obtained Master's degree from  Indian  Institute of Technology, Kanpur and Ph.D. from Indian Institute  of Technology, Madras during 1984 and 1997 respectively.  He is a Fellow of Institution  of Electronics And  Telecommunication Engineers, India and Life Member of Indian Society for Technical Education.

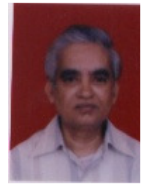

M.Kamaraju, obtained his Bachelor's degree & Master's degree  from Andhra University.   Areas  of interest are Microprocessors, Microcontrollers, Embedded Systems, Low Power VLSI.  He is a fellow of Institution of Electronics and Telecommunication Engineers, India & Member of VSI.

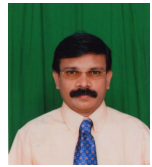